# CHAPTER 3

# DATA STORAGE IN THE DECENTRALIZED WORLD: BLOCKCHAIN AND DERIVATIVES

Enis KARAARSLAN*, Enis KONACAKLI**

*Assistant Professor, Mugla Sitki Kocman University, Department of Computer Engineering, Mugla, Turkey.
E-mail: enis.karaarslan@mu.edu.tr
**Eskisehir Technical University, Department of Computer Engineering, Eskisehir, Turkey.
E-mail: enisk@eskisehir.edu.tr

**Abstract**

We have entered an era where the importance of decentralized solutions has become more obvious. Blockchain technology and its derivatives are distributed ledger technologies that keep the registry of data between peers of a network. This ledger is secured within a successive over looping cryptographic chain. The accomplishment of the Bitcoin cryptocurrency proved that blockchain technology and its derivatives could be used to eliminate intermediaries and provide security for cyberspace. However, there are some challenges in the implementation of blockchain technology. This chapter first explains the concept of blockchain technology and the data that we can store therein. The main advantage of blockchain is the security services that it provides. This section continues by describing these services.. The challenges of blockchain; blockchain anomalies, energy consumption, speed, scalability, interoperability, privacy and cryptology in the age of quantum computing are described. Selected solutions for these challenges are given. Remarkable derivatives of blockchain, which use different solutions (directed acyclic graph, distributed hash table, gossip consensus protocol) to solve some of these challenges are described. Then the data storage in blockchain and evolving data solutions are explained. The comparison of decentralized solutions with the lcentralized database systems is given. A multi-platform interoperable scalable architecture (MPISA) is proposed. In the conclusion we include the evolution assumptions of data storage in a decentralized world.

# 1. Introduction

We are now entering an era where people seek solutions for eliminating intermediaries. The processes can be made faster, while they became less bureaucratic. These solutions can be possible with decentralized solutions; blockchain technology and its derivatives. We mean the "blockchain frameworks" which implement this technology, when we use the term "blockchain technology".

Decentralized solutions are important, as they establish trust without using any intermediary. They do not depend on a central node and are more fault-tolerant and resistant to attacks than traditional solutions. These solutions work as peer-to-peer (P2P), which allows direct communication between peers via the Internet (Karaarslan & Adiguzel, 2018). BitTorrent is one of the most successful implementations of the P2P file-sharing protocol (Alves et al., 2018).

Decentralized solutions can be used to eliminate intermediaries like banks, notary, etc. Bitcoin (BTC) cryptocurrency is a working example of how it can be done. As described in (Brennan et.al, 2018), "cryptocurrencies are only the beginning". Autonomous codes are devised to make the processes autonomous and work without intermediaries. Decentralized applications (Dapp) allow us to have answers within a distributed and secured network (Karaarslan & Adiguzel, 2018).

This chapter aims to describe blockchain technology and to show the differences in its purpose and design. In section 2 we start with a brief explanation of blockchain technology. Blockchain technology fundamentals and security services are described. Data storage in blockchain is addressed here. The challenges of blockchain technology and some remarkable solutions are described in Section 3. Blockchain anomalies, energy consumption, scalability, speed; interoperability, privacy, and cryptology challenges in the age of quantum computing are addressed here. The decentralized derivatives (Tangle, Hashgraph, Holochain) and their technological differences are described in Section 4. Data storage in decentralized systems is covered in Section 5. Evolving data solutions for the decentralized systems and hybrid solutions are given here. Decentralized solutions are compared with centralized databases. A multi-platform interoperable scalable architecture (MPISA) is proposed in Section 6. Finally, results and conclusions are given.

## 2. Blockchain Technology

Blockchain is atechnology, which is used for the keeping of a list of records in a (semi-) decentralized manner. These records contain information about any transaction or any program code (smart contract) which allows a system to work autonomously (Alves et al., 2018; Ali et al., 2017). The records are aggregated in data structures and called blocks. These blocks are linked to each other using cryptographic techniques and thus form a chain structure. The registry, which keeps this blockchain, is called the ledger. Blockchain keeps the ledger distributed and is also called distributed ledger technology (DLT). The ledger is kept in several devices, which are called nodes. These nodes are connected using P2P protocols. These nodes can act as servers or clients at the same time and form a decentralized system. Nodes with different hardware can have different functions, which are summarized in Table 1 (Barnas, 2016). These nodes use consensus protocols to make a common decision on operations, such as the choice regarding who will write the new block. The new block is written by the selected node and then distributed to all nodes.

**Table 1.** Blockchain Node types

| Node Type | Function | Examples |
| --- | --- | --- |
| **Full Node** | Keep full copy of the blockchain, Generate blocks, Validate blocks, Validate transactions, Generate new transaction and broadcast. | Servers or personal computers with sufficient hardware resources |
| **Partial/Half Node** | Keep only partial copy of the blockchain, Validate blocks, Validate transactions, Validate old records as peer support, Generate new transaction and broadcast. | Laptops or alike |
| **Simple Node** | Validate new transactions, Generate new transaction and broadcast. | IoT or limited capacity mobile devices |

Blockchain is not a suitable solution for all computational issues and neither for all data storage problems. The need of a blockchain solution is discussed in detail in Wüst and Gervais' paper (Wüst & Gervais, 2018) and also summarized in Fig. 1. A blockchain solution is suitable under the following conditions:

- If the domain has a dataset which is to be shared with more than one party,
- Where there is low trust between parties and there is no trusted third-party to ensure trust,
- In cases of a need for auditing.

In a scenario of a supply chain, a company may want to track all the processes in the supply chain and even make it transparent to its users. As it is shown in Fig. 2, it becomes complex even in a scenario of two companies (a producer and a consumer). All the transport means, authorities, banks, and others need to share the data or generate transactions during this process. Blockchain is a good solution in a scenario like that, where there are many parties that have to trust each other (Mohan, 2019).

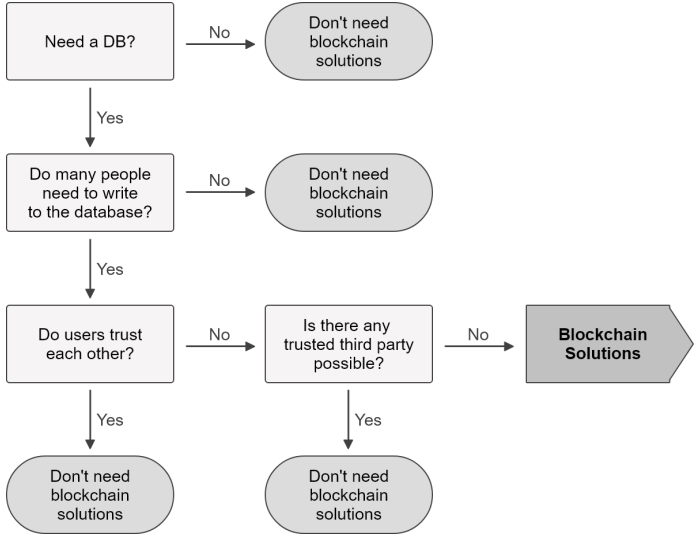

**Figure 1:** Do you need a blockchain?

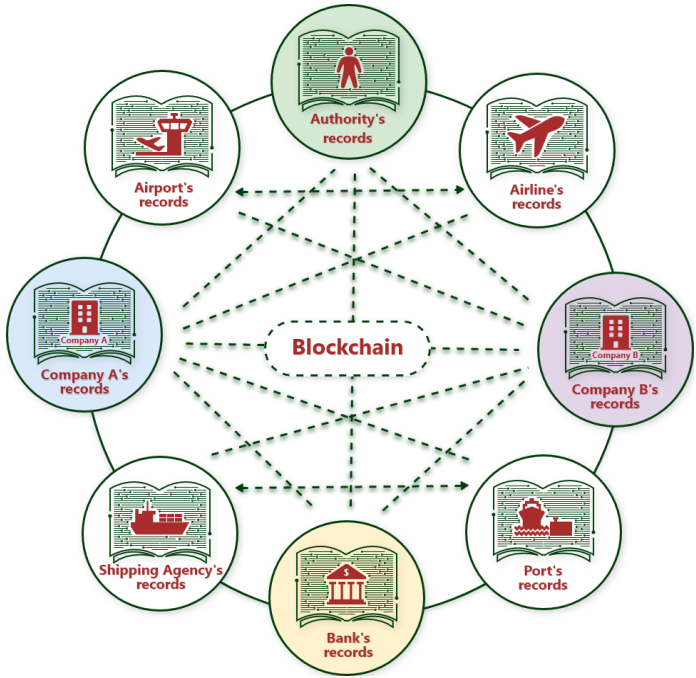

**Figure 2:** Multi-party data access scenario

The blockchain solution will help in keeping the records of the transactions. The records should be reachable at all times, unmodifiable and inerasable (White et al., 2017). The system will work in an autonomous way, which will ensure trust in the system. Full trust, complete privacy, and decentralization should be aimed at when creating such decentralized systems (Karaarslan & Akbaş, 2016).

Different blockchain implementations, which depend on the anonymity and trustworthiness of the validator (node), are possible, as shown in Fig. 3 (Gür et al., 2019). These are:

- Allowing nodes to join the network with or without permission (permissionless),
- Allowing public or private access to the ledger,

Different consensus protocols such as proof of work (PoW), proof of stake (PoS), proof of authority (PoA), practical byzantine fault tolerance (PBFT) and such like are preferred in accordance with the anonymity and trustworthiness of the node.

There are also hybrid blockchain implementations, which allow different types to work together to achieve a function. Some implementations can have public and private ledgers together. Implementations like federated (consortium) blockchain allow multiple organizations to share information privately between parties (Bauer, 2015).

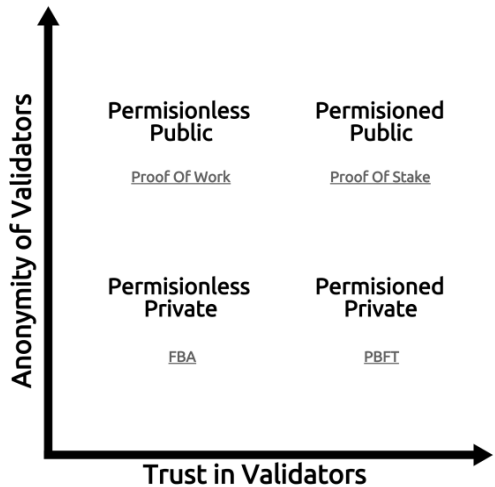

**Figure 3:** Blockchain implementation types per anonymity/trust of validators

Cryptocurrency implementations mostly use permissionless validators and public blockchain. The users are anonymous or pseudonymous. The term 'permissionless' states that any node can enter or leave the system without permission. The trust in the validator is low as the nodes are anonymous (Karaarslan & Akbaş, 2016). Cryptocurrency implementations depend on using their own currencies to run transactions on their systems. Bitcoin can be called the first blockchain implementation of this type, which has been active since 2008. Bitcoin is the proof-of-concept that this type of system can work and have value. Satoshi has proposed a model in his paper (Nakamoto, 2008), where the system generates a new crypto coin per block and gives an award to the owner of the node that will write the block. Ethereum (ETC) introduced a framework where new blockchain applications can be developed. Smart contracts are used which are in the form of an autonomous software code on the blockchain.

The steps in the process of making a value (cryptocurrency) transfer in such a blockchain network are given in Fig. 4. In this scenario, Fatih wants to make a value (cryptocurrency) transfer to Eylul. Most cryptocurrency systems use "mining pools", which orchestrate such a process. The nodes in the P2P network validate the transactions (account balance check, double spending check) and collect the validated transaction data. According to the protocol used, nodes collect information of variable number of different transactions in a specified time. PoW consensus protocol is used to select the node which will write the new block. PoW depends on a calculation to solve a puzzle like a mathematical problem. The node, which solves the problem, will first be selected. The selected node will form and write the block and advertise it in the network (Karaarslan & Akbaş, 2016). The fairness of the node selection and the security of such a process results in high energy and time-consuming operations (Gür et al., 2019). PoW and alike consensus protocols, have a bad reputation on high energy usage, which is said to affect climate change. The blocks are transparent and that means the transaction information is visible through web interfaces, which are called explorers such as the block explorer. These web interfaces also show detailed information about that cryptocurrency system.

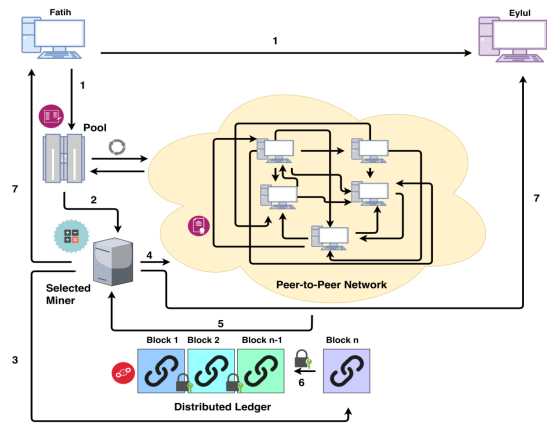

**Figure 4:** Transaction steps of a value transfer in cryptocurrency implementations

Different consensus protocols can be developed and deployed, which consume less energy and are faster than PoW. POS and alike consensus protocols are being tested with cryptocurrency implementations (Zheng et al., 2018). This will be covered in section 3.2.

The needs of enterprise implementations are different from cryptocurrency implementations. The identity of the users is known. Permissioned validators and private/public blockchains are mostly used. Different parties of the blockchain system supply the validator nodes. The validators are trusted and not anonymous, which means they are under the control of the management. PoW consensus protocol is not necessary. PBFT, PoA and similar consensus protocols are preferred in this type of implementation (Zheng et al., 2018). Hyperledger Fabric, R3 Corda (Valenta & Sandner, 2017) and Quorum can be given as examples.

Hyperledger Fabric is widely used in production (Hyperledger, 2018) and in academia (Androulaki et.al, 2018; Nasir et al, 2018). Such an implementation scenario in Hyperledger Fabric is given in Fig. 5, which consists of a customer and his/her IoT device, two companies, and one authority. The customer can be subscribed to different companies and there is also one authority that these companies have to share their data with. The IoT device of the customer sends a summary of collected data to the blockchain network. Each company creates a group (channel) among themselves. Different consensus protocols and different types of nodes can be used in each group. These nodes, rest server, and CA server can all be installed as Docker containers. Each node has limited authority. The owner of each transaction is identified in its own certificate authority. These groups also share data with the authority, which is labeled as Auth. C Peer in this case (Gür et al., 2019).

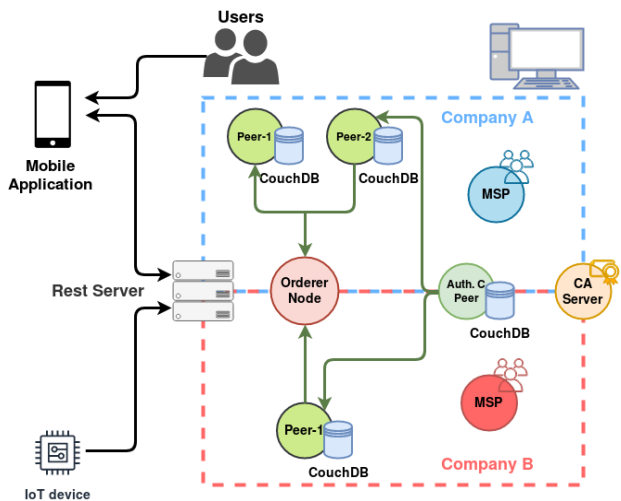

**Figure 5:** Hyperledger Fabric enterprise blockchain solution

Many cloud services have started to provide environments of blockchain as a service (BaaS), which serve cloud services to build Dapps. By way of example, IBM, Alibaba, Huawei, and many others provide Hyperledger Fabric based BaaS (Mohan, 2019).

A comparison of the security services provided by blockchains, central databases and distributed databases are given in Table 2 (Bozic et al., 2016). Data integrity, availability and fault tolerance services can best be provided with blockchain.

The integrity of the data is established by the design of the DLT. Each block is connected to the previous one using its hash value. Hash functions (SHA-256, Keccak-256 … etc) are one-way functions that form the fingerprint of the input data. This data structure makes the binding so strong that, when an attacker wants to change block n, the blocks starting from the nth block, until the last block, have to be modified and rewritten according to the change. It should also be noted that the selected node of each block is also recorded in the ledger and such an attempted attack will easily be detected (Karaarslan & Akbaş, 2016).

The availability and security of the system depends on the number of nodes and their distribution in the network. More nodes will make the system stronger against the attacks. Then taking control of the majority of the nodes by the attacker will be harder and the compromised nodes will not be able to misguide the block creation process. If the nodes are more widely distributed in different networks, the network will also be stronger against DDoS attacks. Fault tolerance is the ability of the blockchain to correct any misuse and errors. This is implemented by using consensus protocols.

Privacy is not a design concern in most implementations such as cryptocurrencies. Privacy and security in Bitcoin are investigated in (Conti et al., 2018). Different implementations can have different levels of privacy. Transparent records do not mean the privacy level is low. Personal data is not revealed, transactions are only traceable with the public addresses. Transparency property is used to prevent any possible fraud and misuse. It can be used to enable safer environments (Ölmez & Karaarslan, 2019).

| Table 2. Comparison of the security services | | | |
|---|---|---|---|
| | *Blockchain* | *Central Database* | *Distributed Database* |
| **Integrity** | High | Average | Average |
| **Availability** | High | Low | Average |
| **Fault tolerance** | High | Low | High |
| **Privacy** | Variable* | High | Average |
| *\* Privacy is not by design. Mainly depends on the implementation* | | | |

There are nascent standardization efforts, which focus on narrow aspects of blockchain (Mohan, 2019). IEEE Blockchain Initiative has just started several blockchain standardization efforts focusing on areas like agriculture, medicine and IoT (IEEE, 2019). ISO/TC 307 technical committee is working on blockchain and distributed ledger technologies (ISO, 2019). W3C community group is working on the Web Ledger Protocol, which will describe the format and protocol of decentralized ledgers on the web (W3C, 2019).

### 3. Meeting the Challenges

Despite the opportunities of blockchain technology, the challenges of blockchain are still notable for discussion. The challenges can be summarized as follows:

- Blockchain anomalies,
- Energy consumption,
- Scalability and speed,
- Interoperability,
- Privacy,
- Cryptology challenges in the age of quantum computing.

Some note-worthy solutions proposed and studied are given in the subsections.

### 3.1. Blockchain Anomalies

Some anomalies may result in the addition of conflicting blocks and the formation of new branches of the chain in PoW based blockchains. The conditions, which may lead to these anomalies, are covered in Natoli and Gramoli's study (Natoli & Gramoli, 2016). This can cause usability, integrity and performance problems (Mohan, 2019). Blockchain implementations should give deterministic guarantees on these conditions. Implementations can be adapted and smart contracts can be written to overcome these types of anomaly (Natoli & Gramoli, 2016).

### 3.2. Energy Consumption

Mining operations of the conventional PoW based blockchain systems require expensive hardware and a very high degree of energy consumption (Flipo & Berne, 2017; Trautman & Molesky, 2019). Energy efficient solutions, which will replace or minimize the usage of the conventional PoW based blockchain systems, are being experimented. Different node selection algorithms are proposed which are based on random choice or on the cryptocurrency amount of the miners (Rosic, 2017).

POS consensus protocol has started to be preferred in cryptocurrency implementations instead of the PoW approach. The nodes have to deposit a predefined amount of cryptocurrency and show their commitment to the system and become a trusted validator. The system does not require a calculation-based competition, rather it randomly chooses from the validators. The possibility of being selected is directly proportional to the amount of cryptocurrency. The system will consume much less electricity and be much faster with POS (Sayeed & Marco-Gisbert, 2018; Opray, 2017).

Current business blockchain frameworks such as Hyperledger and R3 Corda are token-free platforms and are far more energy efficient as they eliminate this extravagant process. Other blockchain derivatives, such as Hashgraph, Holochain, and Tangle, are also energy efficient and resource friendly DLT systems.

### 3.3. Scalability and speed

Scalability is the ability to handle large volumes of transactions at high speeds. This basically depends on the following factors:

- Consensus: The nodes have to agree on the validity of the transaction. Adding information to a block with POW consensus protocol is a very slow process in the conventional cryptocurrency architectures. Creating a block can take around 10 to 60 minutes in Bitcoin (Bitinfocharts, 2019); it takes about 15 seconds in Ethereum (Etherscan, 2019). All new blocks are broadcasted and verified by all nodes in a typical blockchain network.

- Storage: Storage capacity is the biggest concern when implementing blockchain. The exponential growth of the block size creates a performance problem. Keeping the whole data in every node can be unfeasible and impractical in many solutions.

This brings out the scalability problem since the broadcast traffic and the size of the ledger data stored in the nodes increases exponentially because of the nature of the blockchain architecture. Moreover, lightweight devices like Internet of Things (IoT) do not have sufficient resources for this. Many solutions have started to use all nodes for validation of the transactions, and only use some (full nodes) for storing all the data. Maintaining only the summary or link of the data in the nodes, keeping the data in the DSN architecture is also being implemented. Vitalik Buterin, co-founder of Ethereum, once claimed that a blockchain solution can have a maximum of two characteristics out of the three core characteristics (decentralization, security and scalability). This is also called the scalability/blockchain trilemma, which is shown in Fig. 6. An attempt to solve the scalability problem will result in sacrificing on decentralization or security (Gomez, M., 2017).

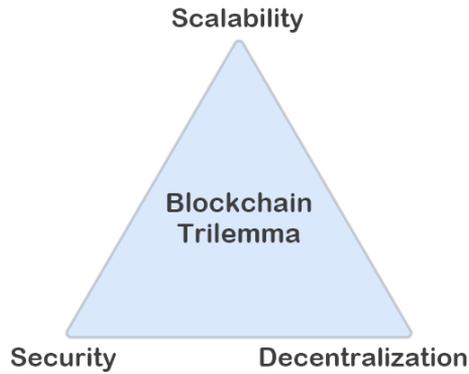
**Figure 6:** Scalability/Blockchain Trilemma

Scalability solutions can be covered in four layers; hardware, network, blockchain and application (EUBlockchain, 2019a). These solutions are summarized in Table 3. However, the throughput values are estimations to show the effect of each solution.

Using better machines or having faster communication has a limited effect on scalability. Hardware upgrades can be in limited amounts, and this type can perform best only in consortium chains and dPoS consensus protocol (EUBlockchain, 2019a). Higher bandwidths may be available from the telecom providers, but that does not mean that faster communication is possible.

There are various scalability solutions in blockchain layer such as adjusting block size, adjusting block interval, sharding, using different consensus mechanisms and decentralized derivatives. Solutions like directed acyclic graph (DAG), distributed hash table (DHT) can also be used. These solutions have different requirements and are difficult to compare with each other on performance (EUBlockchain, 2019a). These will be covered in detail in the next sections.

**Table 3.** Scalability measures in blockchain layers

| Layer | Solution | Throughput | Limitations |
|---|---|---|---|
| Hardware | Using better machines | Up to 5-10x | Not for large networks<br>Best only in consortium chains, dPoS |
| Network | Faster communication links | Up to 5x | Not affordable in all areas |
| Blockchain | Adjust block size<br>Adjust block interval<br>Sharding<br>Different consensus mechanisms<br>Decentralized derivatives | Up to 10-20x | Difficult to compare on performance |
| Application | Off-chain<br>Sidechain | Up to 10,000 to 100,000x | Depends on interoperable tools |

Significant parts of the data and computation can be transferred to conventional systems to make the processes faster. Structures like off-chain and side chain can be used to increase the throughput. Direct channels can be established between parties (EUBlockchain, 2019a). Solutions will be described in detail in the next sections.

### 3.4. Interoperability

Interoperability of blockchain infrastructures has emerged as a newborn challenge for the blockchain community in recent years. Although blockchain technology has been designed and established for removing the intermediaries and trusted third parties, users of different blockchain systems cannot easily transfer digital assets between each other without using an intermediary. For example, if a user wants to transact some data or a digital asset, secured and processed in Hyperledger Fabric network, to a R3 Corda network client, this user first has to register to the Hyperledger Fabric network, then decrypt the secured data, and then register on R3 Corda to use this network's functionality and put the aforementioned data into R3 Corda network. This creates a great amount of wasted time and processes. It becomes a necessity to ensure the interoperability of different blockchain architectures even between different companies or industries.

We will testify that different blockchain architectures will be able to communicate and share digital assets in the near future. Mechanisms like QuickX should be used to enable cross transactions. Sidechains have been proposed as a promising mechanism that allows transactions from one blockchain to another. It is not only a DLT technology but also a potential architecture for enabling the interoperability of the blockchain technologies (Ray, 2018).

### 3.5. Privacy

Privacy is another challenging issue that emerges from the nature of the blockchain methodology. In a permissionless blockchain architecture, all parties have the right to download the ledger, which implies that they have the right to explore the entire history of the recorded transactions. Implementing "the right of privacy" is a challenge in these architectures. Special care must be taken, when working with PII (Personally Identifiable Information). It is a good practice not to store PII on the blockchain and let the user handle his/her own data.

Zero Knowledge Proof (ZKP) can be integrated into blockchain systems to ensure privacy. The user can be given the total control of his/her data. ZKP can be used to validate any process (like identity check) without revealing any information about it (Goldreich, 2019; Korkmaz et al., 2019).

### 3.6. Cryptology challenges in the age of Quantum Computing

Quantum computing and the parallel processing power it promises threaten the security of the current public-key-based algorithms and blockchain systems. Quantum computing is an earthshaking technology that can be used to break ciphers and expose secrets that are secured by the current cryptographic algorithms (Piscini et al., 2018). Symmetric algorithms appear to be secure against quantum computers (and Grover's algorithms) by simply increasing the associated key sizes. Commonly used public-key cryptographic algorithms (based on integer factorization and discrete log problem) such as RSA, DSA, Diffie-Hellman Key Exchange, ECC, ECDSA will be vulnerable to Shors algorithm and will no longer be secure (Cromwell, 2015).

Researchers are studying post-quantum blockchain (PQB) and secure cryptocurrency schemes based on PQB systems, which can resist quantum computing attacks. This area is still under progress (Gao et al., 2019).

### 4. Decentralized Derivatives

There are Blockchain derivatives that intend to solve the problematic issues of this technology and offer individual solutions for specific aforementioned challenges (Schueffel, 2017). These derivatives are basically distributed ledger technologies that have different consensus protocols and architectures other than conventional blockchains. Directed acyclic graph (DAG) and distributed hash table (DHT) aim to perform the benefits of blockchain with better performance. Sidechain implementations offer to solve scalability solutions. Gossip protocol aims to reach a faster consensus than the counterparts do. These solutions are described, then the platforms that use these solutions are compared.

### 4.1. Directed Acyclic Graph

Changing the manner of the transaction validation process using distributed acyclic graphs is a new and effective approach, which creates new solutions for the scalability and speed problems of traditional blockchains. IOTA Tangle and Byteball are well-known examples, which put this methodology into practice (Wang et al., 2018).

The graphs are a representation of the connected peers through which information can be passed from one peer to another along different edges in a multidimensional space. They are great tools for traversing between various connections of individual units of data. A peer initially communicates with the closest peer according to pre-defined rules. They may be directed or undirected. Fig. 7 shows various graph types.

DAG is a non-looping graph that joins edges to turn in a pre-defined direction. Each square stands for a separate transaction in Fig. 7. The transactions are validated by the recently validated transactions in the way through the DAG branches.

DAGs stand out as promising DLT structure, enabling promising applications that can compete with classical blockchains (Jiab, Bouric, Guntad, & Roubaude). The use of DAG structures in distributed networks aims to solve the speed, cost, and scalability challenges of classical blockchain architecture.

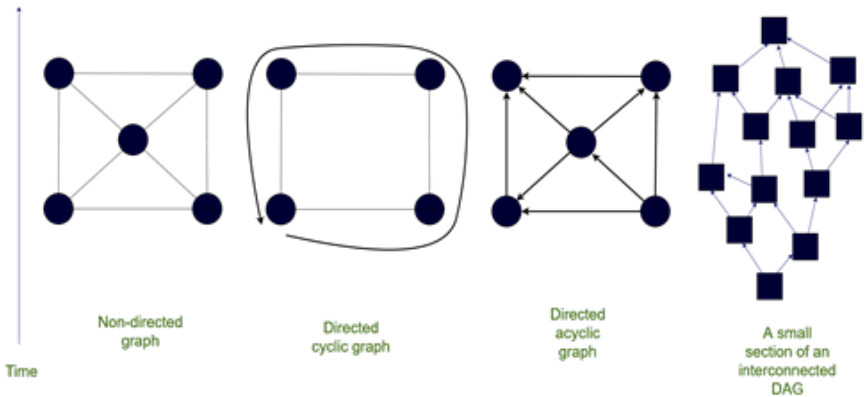

**Figure 7:** DAG and other graph types

### 4.2. Distributed Hash Table

DHT is a set of distributed storage systems that provides lookup and storage schemes for the peers, which store and retrieve data, identified by key values in the network. Distributed hash table establishes a distributed routing table in a very large and distributed network. There is no central authority, and peers can join and leave the network at any time in a distributed network. They are connected together through an overlay network. The nodes store and share the data by coordinating with each other (Dufel, 2017).

Fig. 8 shows the dictionary-like structure of the DHT usage. DHT allows the nodes to find any given key in the key-space. It maps the whole network by key values. Key value is the ID of the node that is calculated by hashing the node's IP and port combinations. This key identifies every separate node, and the node's position in the DHT indicates separate independent node which keeps related data. If a node leaves the network, the algorithm automatically shifts the abandoned key value to another peer, which is not addressed with any key. Nodes can make a search for the related node and find its data using this easy-to-implement structure. (Dufel, 2017).

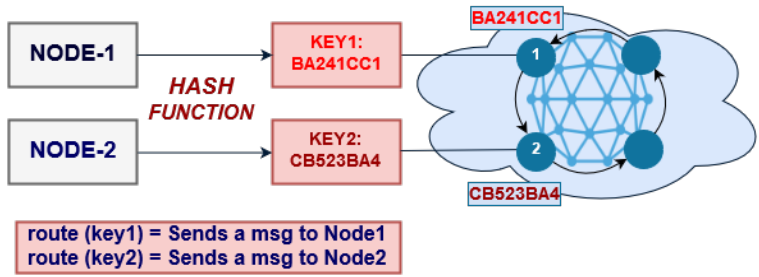

**Figure 8:** Mapping network using key values

### 4.3. Side Chain

Sidechain (child-chain) is a solution which allows making a partial copy and a separate branch of the (main/parent) blockchain, which is bound to counterpart(s). The original blockchain is called the mainchain and all additional blockchains are called the sidechains. Sidechains are used to allow cryptocurrencies and other digital assets to be processed in a separate private blockchain and then be securely transferred back to the original blockchain (Halpin & Piekarska, 2017).

Sidechain uses two-way pegging mechanisms to allow two separate chains bound to each other and transfer assets in between. In a crypto currency transfer scenario which is shown in Figure 9, a user on the parent chain initially sends its cryptocoins to an output address (Musungate et.al, 2019). The first step is a lock box, which locks the sent cryptocoins so the user cannot spend them. After acceptance of the transaction, an equivalent amount of crypto coins is delivered to the side chain. The user can spend the coins after that step. The reverse process is performed when moving back from a sidechain to the mainchain.

Every sidechain is responsible for its own security. Since each sidechain is independent, if it is hacked, the damage will be enclosed within that chain and will not affect the main chain.

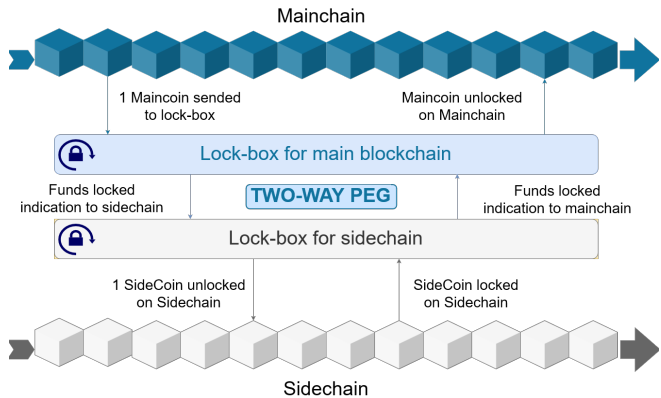

**Figure 9:** Transaction process of sidechain and two-way pegging

### 4.4. Gossip Consensus Protocol

Gossip is a communication protocol that is an agent for nodes to interact each other at high speed (Baird et al., 2018). When a node gossips, it randomly selects a peer and shares the received new information with it. The selected peer does the same thing, and this process continues until the information is passed on to all the connected nodes. It works just like social gossiping and information passes through the whole network in this way. The transactions are validated by the previous successful transactions rather than the mining process of currency based blockchain architectures. This protocol can run successfully on DAG and DHT networks to achieve high speed transactions (Zhenyu, Gaogang, Zhongcheng, Yunfei, & Xiaodong, 2018).

### 4.5. Comparison of the DLT Technologies

The DLT systems covered in this section are typically designed to deal with a registry of data that is distributed across a network. They are more transparent and robust compared with the conventional centralized database systems. The basic idea of blockchain derivatives is to form alternative decentralized systems that can solve the structural challenges and overcome the architectural limitations of traditional blockchains. Table 4 shows the comparison of the DLT derivatives.

Even though there are many similarities among the derivatives of DLT systems, there are also some architectural differences (El Ioini & Pahl, 2018). The foremost blockchain designs were created to be permissionless, but the DLT derivatives are predominantly permissioned. Everyone can join a permissionless network, however only accepted parties may access the network in the permissioned DLT alternatives. This difference also influences the size of the network. Blockchain networks, which are used for well-known cryptocurrencies such as Bitcoin, aim to expand to provide a more secure environment. In a permissioned DLT network, the number of parties involved tend to be smaller, as this number does not have such an effect on the security of the systemin this case.

Projects such as Hashgraph, Holochain and Tangle are promising platforms, which are creating new types of distributed ledger technologies. These projects share the common aspects of distributed, consensus, flexible, and peer-to-peer platforms. Although being fast DLT architectures, they use their own consensus protocols and the data structures. Hashgraph and Tangle solve the scalability problem with DAG.

Hashgraph and Holochain use the gossip consensus protocol. Hashgraph is a patented permissioned DLT network, which can handle over 250,000 TPS. The validation of the network requires at least ⅔ of the nodes to receive gossip and confirm transactions. Holochain uses distributed hash table (Anwar, 2018). It aims to create a new distributed Internet structure, trying to establish a new secure generation of cloud computing framework. Trust is established using the computing power of the peers. It is estimated to have an immense scalability rate depending on the holochain networks expansion. Each user peer keeps its own data and transactions information (Harris-Braun, Luck.,& Brock, 2018).

Bitcoin processes 3–7 transactions per second (TPS) and Ethereum can handle 10–20 TPS, however Hashgraph promises to process hundreds of thousands of TPS (Kerner, 2018). Tangle's consensus mechanism hashcash has a high theoretical limit on the TPS throughputs (IOTA, 2019). Tangle can reach up to 800 TPS rates (Kerner, 2018). Despite their potential advantages, their capabilities have not been tested as traditional blockchain systems.

Sidechain has a very high potential to enhance scalability and TPS values depend on the platform used. It can also be used to provide interoperability between different blockchains. There are several platforms which are testing sidechains. A promising project is Plasma (Saini, 2018). Plasma is the child-chain solution of Ethereum. High TPS values are aimed at by allowing each Dapp to use its own chain (Poon, & Buterin, 2017).

**Table 4.** Comparison of the DLT Technologies

|  | *Blockchain* | Plasma | *Tangle* | *Hashgraph* | *Holochain* |
|---|---|---|---|---|---|
| **Structure** | P2P | P2P | Directed acyclic graph | Directed acyclic graph | Distributed hash table |
| **Platform** | Bitcoin | Sidechain | IOTA | Hedra Swirlds | Holo |
| **Transaction per second (Tps)** | 4 to 7 | More than billions | 500 to 800 | More than 200.000 | More than millions |
| **Consensus** | PoW | POS | PoW: hashcash | Virtual voting | DNS Validation Rules |
| **Decentralized** | Yes but using mining pools make it semi decentralized | Depends on the implementation | Semi-Centralized | Semi-Centralized | Decentralized |
| **Licence** | Open Source | Open Source | Open Source | Patented | Open Source |
| **Maturity** | Proven and been used since 2008 | Experimental | Experimental | Experimental (Public use since 2018) | Experimental (Alpha1 rel. in 2018) |

**5. Data Storage in Decentralized Systems**

Blockchain is not a place to store all kinds of different data. As mentioned above, it is a registry where the records (logs) of the transactions are kept. A transaction can be a record of any process and may also contain codes, which allows the autonomous working of a system. A transaction can also give a link to the cloud storage where the actual data exits. Data may be stored in different forms. There are evolving data solutions to solve scalability and interoperability problems. Selected solution proposals are covered first. Then the hybrid solutions, which are formed by using different solutions together, are covered. This section will continue with the comparison of decentralized solutions with centralized databases.

**Table 5.** Comparison of Blockchain with Evolving Data Solutions

| Category | Solution | Throughput | Cost Power& Resource | Capacity Block width/size | Advantage | Disadvantage |
|---|---|---|---|---|---|---|
| Basic Blockchain (Bitcoin) | PoW | Low | High | Basic | Proved to work in trustless environment, Protection against DDoS attacks | Scalability, Computationally Expensive, Needs high computational power, High energy and processing costs, %51 Attack |
| Alternate Consensus Protocols | POS PoA | High | High | Low | APX 0 runtime cost, High transaction speeds | The node who has the steak controls the network |
| | Raft-based consensus | High | Low | Low | Handle multiple problems, Easy to implement | Lacking enough live tests |
| | Gossip | High | Low | Low | APX 0 transaction fee and waiting time | Lacking enough live tests |
| On-chain | Big block | High | Low | High | High capacity transmissions | Centralization of mining pools, High Orphan block rate |
| | Segwit | High | Low | - | Various Possible Bitcoin solutions | Fungibility occurrence |
| | Sharding | High | - | Low | Low capacity burden Parallel processing | %1 Attack |
| Off-chain | Lightning network | High | Low | Low | APX 0 transaction fee and waiting time | P2P Payment channels |
| | Raiden network | High | Low | Low | General purpose channel | P2P Payment channels |
| Child-chain | Plasma | High | - | Low | Parent-child blockchain tree | High costs of verification |
| Inter-chain | Side Chain | High | Low | Low | Blockchain interoperability and cross transactions | Application boundaries |
| DSN | IPFS Gaia Storj | High | Low | High | More secure Flexible Reduced rate of data failures and outages | - |
| Decentralized Derivatives New Solutions | DAG (Nano, IOTA & Byteball) | High | Low | Low | Better scalability No miners Quantum resistant cryptography | Lacking enough live tests |
| | DHT | High | Low | Low | APX 0 runtime cost, Very high transaction speeds | Lacking enough live tests |

## 5.1. Evolving Data Solutions

Data storage solutions are evolved to solve the scalability, interoperability, and privacy of blockchain in several forms and are given in Table 5. The solutions are shown as follows (Kim et.al, 2018):

- Using alternative consensus protocols

- On-chain: Storing all the data on the main-chain. Solutions such as sharding, making blocks bigger are possible.

- Off-chain: Off-chain is storing the data outside the blockchain, processing it and writing the summary on the blockchain. There are challenges to reach the manipulation resistance, verifiability and privacy (Eberhardt & Tai, 2017), (Lightning network, Raiden network).

- Child & parent chains: The records of the child-chain are processed and written to the parent-chain in this type of implementation (Plasma).

- Inter-chain: This is used to provide communication and join the functionality of different blockchains. Structures such as Atomic swaps and side-chain are used.

- DSN: Using blockchain and cloud storage together.

- Other structures (DAG, DHT … etc)

Enterprise solutions have different needs and expectations than the cryptocurrency systems. Different consensus protocols (PoA, POS, raft-based consensus, Istanbul BFT, etc.) are being implemented to make the consensus phase faster and reach higher transactions per second (TPS) rates by eliminating the mining processes, while ensuring confidentiality. By way of example, Quorum, which is based on Ethereum, does not use POW/POS consensus protocols, but instead supports multiple consensus protocols to support enterprise needs. It supports alternative consensus protocols like raft-based consensus, Istanbul BFT (IBFT) (Baliga et.al. 2018).

Sharding and making blocks bigger are both possible on-chain solutions. The Big block is the basic process to enhance the block size. Making blocks bigger enlarges the transmission limit, but big blocks need extremely high processing powers, which will also increase the transmission cost (Clifford, 2017). Since the propagation speed becomes limited, this process increases the probability of orphan blocks appearing. Big blocks are not efficient at the current stage because of these disadvantages. Sharding is the process of dividing a database into smaller segments. It is also called horizontal partitioning. Sharding is a controversial issue in blockchain and there are different views. Vitalik Buterin and Beniamin Mincu believe that sharding can be one of the solutions. Vitalik Buterin once expressed the concept of the 'sharding' model as the creation of hundreds of different universes, each of which being different account spaces. According to him, the transaction will affect the things only in the universe it belongs to. He claims that thousands of transactions per second can be achieved without any special server, nor with consortium chains (Gomez, M., 2017). Beniamin Mincu, the CEO of the Elrond Network, claims that sharding is needed to reach the throughput capacity that is needed to rival networks like VISA and states that some challenges are single-shard takeovers, cross-shard communication and data validity (Cointelegraph, 2019).

Lightning network and raiden network can be given as examples of off-chain solutions. A consensus process will not be used in lightning network when two parts trust each other. Transactions will be quicker and will not be recorded on the chain (Karaarslan & Adigüzel, 2018; Poon & Dryja, 2016).

Plasma can be given as an example of the child-chain solution in Ethereum. Each Dapp will use its own chain in the Plasma solution (Poon, & Buterin, 2017).

Atomic swaps and sidechain are inter-chain solutions that are established to enable cross transactions and blockchain interoperability. Atomic Swap is the peer-to-peer currency exchange between different blockchain networks, without the need for a mediator. Sidechain was covered in Section 4. It has a very high potential of enhancing scalability and can also be used to provide interoperability between two separate blockchains (Musungate et.al, 2019).

Using blockchain and cloud storage together forms decentralized storage, which is also called a decentralized cloud storage network (DSN). Data can be stored and shared without having to trust third parties (Wilkinson et.al, 2014). This solution is used to overcome storage limits and also provide personal data storage and privacy. A DSN network can have advanced privacy, security and data control as it has the following characteristics (Karaarslan & Adiguzel, 2018):

- More secure as it uses client-side encryption,
- Flexible as there are speed and low-cost advantages with proper implementation,
- Integrity and availability of the data is ensured with proof of retrievability,
- Reduced rate of data failures and outages.

Examples of DSN can be given as follows (Karaarslan & Adiguzel, 2018):

- Gaia: It is used by Blockstack. When the user uses the decentralized application and any data is needed to be written, it serves to save this data on the existing cloud infrastructure. Data is written in encrypted or signed form (Ali et.al, 2017).
- Storj: Storj works as a P2P cloud storage network. An open source software project called Metadisk provides a set of tools to make Storj easily integrated with legacy systems (Wilkinson et.al, 2014).

Other token-free DLT derivatives, such as Hashgraph, Holochain, and Tangle achieve better scalability and TPS rates by using different structures (DAG, DHT) while eliminating mining operations.

Different multi platform solutions are possible. Outstanding ones are shown as follows:

- Using hybrid blockchain solutions which involve public and private blockchain solutions working together,
- Using inter-chain structures like sidechain to make different decentralized solutions working together.
- Using blockchain with decentralized cloud storage network (DSN)
- Alternative cloud storage platforms which use blockchain as an awarding system.

### 5.2. Comparison of Decentralized Solutions with Centralized Databases

The differences between decentralized solutions and databases is in their design and purpose. This topic is widely investigated in (Tabora V., 2018). Blockchains are distributed systems, which hold replicated databases on several different nodes. Special consensus protocols are used to ensure these replicas are trusted (Murthy C., 2016).

Database systems are becoming more complex with the ever-increasing usage of different data types, big data, and cloud infrastructure. There are many characteristics to classify them. Firstly, it is important to talk about the data management models like relational and non-relational. Relational databases are the most commonly used database types in the world. However, non-relational databases are also becoming popular with the rising storage needs of unstructured data and the increasing usage of the machine learning processes which use them. By way of example, No-SQL databases are also becoming widespread and are mostly used for rapid development or used to store large amounts of data that have little or no structure. Blockchain is a non-relational database but there are also exceptions. A recent blockchain system called postchain (Botsford A., 2019) seems to be the first blockchain system which uses the relational model.

The general characteristics of decentralized solutions with relational databases are given in Table 6. There are also variations but it is outside the scope of this section. The comparison of these solutions is shown in Table 6 and is summarized in the following paragraphs.

Database is deployed in client/server model, however blockchain system is decentralized. There is mostly one party involved in relational database. Consistency is hard and expensive to achieve in relational database when there is more than one party. Blockchain solutions are best suited for multi-party solutions and satisfy consistency as all nodes have the full copy of the dataset. A blockchain system will directly identify and correct possible inaccurate records. Companies, authorities, banks, transportation companies and such like can be a part of this multi-party network (Schlapkohl, 2019).

Security services such as availability, integrity, and fault tolerance are highly supported with blockchain systems. Database systems may be deployed to serve these services, but we can say that it will not be as effective as blockchain systems. Users trust databases that they will work right, but no one can be sure since administrators have full control of the system. Even competitor companies need to share data between each other. They do not need to trust each other, but need to trust the shared data. Trusted third parties can also be used to ensure trust but their trustworthiness is also questionable (Karaarslan & Adiguzel, 2018). Blockchain systems work by ensuring trust without using any intermediaries. Trust is established using autonomous code and consensus protocols.

The attackers try to delete all possible evidence on the compromised system after any attack. Digital forensics become difficult when logs are deleted. Any change attempt on the blockchain ledger is also kept on the immutable ledger, so the details of the incident (who, when and what) will be detected. This will also have a deterrent effect on attackers.

Cryptocurrencies use public chains that have transaction records transparent to everyone. They allow everyone to see and query all transaction records on the system. Enterprise solutions use private or hybrid chains and queries that give reading access only in that domain. Databases do not give such a service.

Data management is relational in databases, blockchain is non-relational. The user accounts are created on the database system and administered. Security is mostly implemented by giving roles on the database system such as the database tables they can reach and their permissions. However, blockchain works autonomously; consensus protocols and smart codes (autonomous codes) define how the system works. There are no users on the system. Decentralized identity management systems (IDMS) (EUBlockchain, 2019b) or such like may be used, however these do not define any user roles on the system. Permissioned blockchains are also possible where there is an access control layer. This layer is used to permit specified actions to the defined users. This property is different from the relational database permission process.

Blockchain is distributed by default. Database systems are installed as standalone by default, but may also be deployed asdistributed. However, the amount of nodes that the blockchain solutions can reach is mostly not possible in distributed database solutions. Only allowed nodes can be added to the distributed architecture of relational databases. Nodes can be permissioned or permissionless depending on which decentralized technology is used. Redundancy is only possible to the level where relational database is distributed. All full nodes have the latest copy and data redundancy is satisfied in the blockchain implementations.

Sharding is available in relational databases when the data is distributed in several servers. Sharding is a controversial issue in blockchain.

Parallelization is limited in relational databases. Cloud adaptability is high with decentralized databases. Big data handling capability is limited in relational databases; however, decentralized solutions are more suitable for big data operations, especially when used along with the cloud infrastructure. Relational databases generally handle small data better. However, decentralized solutions handle big data better. Scalability for variable data sizes is rigid in relational databases, but elastic in decentralized solutions (Demir et.al, 2018).

Databases support high volume transactions at a fast processing rate. Blockchain implementations have to validate transactions and this comes at the cost of speed. The solutions which use PoW consensus protocols support low volume transactions at a slow processing rate. Higher volume transactions and faster processing rates are possible when different consensus protocols like PoS, PoA are used. These faster consensus protocols mostly need trusted nodes. We can say that blockchain should not be used when transaction speed is a concern. However, there are studies on low-latency solutions. Data analytics is supported with databases, however blockchain can be described as poorly supported in this concept.

Blockchain systems are said to have problems in the areas of data size, synchronization, energy consumption, interoperability and scalability. There are many studies and many new solution proposals on these areas. Some of these proposals are given in the previous sections of this chapter. Databases are widely used in various projects. However blockchain solutions have a value when there is a need for establishing trust between parties without any intermediaries involved and a need for data verification.

| Table 6. Comparison of Blockchain (and derivatives) with Relational Database | | |
|---|---|---|
| | **Relational Database** | **Blockchain (and Derivatives)** |
| Centralization | Centralized | Decentralized |
| Party Involved | Mostly one | More than one party |
| Consistency (multiple party) | Hard and expensive to achieve | Consistent (full copy) |
| Security Services (Availability, Integrity, Fault Tolerance) | Poorly supported (by default) | Highly supported |
| Trust | Trusted 3rd party | Trust without intermediary Trust on smart code, consensus |
| Forensics | Difficult (if logs are deleted) | Easier (unalterable records) |
| Transparency of transaction data | No | Yes (public chains) Partial (private or federated chains) |
| Data management system | Relational model | Non-relational |
| Management Method | Administrated | Autonomous |
| User Control Method | Permissioned | Permissionless, permissioned |
| Distributed Deployment | Possible | Distributed by default |
| Node Add Method | Permissioned | Permissionless, permissioned |
| Redundancy | Possible (when distributed) | All full nodes have the latest copy |
| Sharding | Suitable (when distributed) | Controversial |
| Parallelization | Limited | Suitable |
| Cloud Adaptability | Limited | High |
| Big data handling | Limited | Suitable (with cloud) |
| Scalability for variable data sizes | Rigid | Elastic |
| Read/write Speeds | Faster for small data | Faster for big data |
| Transaction Volume | High volume | Low to Average* |
| Transaction Speed | Fast | Slow to Average* |
| Data Analytics | Supported | Limited |
| Problems | Single Point of Failure Administration Issues Security Issues | Energy Consumption Interoperability Scalability |
| Best When | High volume of data Fast processing need Quick query need | Data verification needed Establishing Trust without intermediaries |
| *Changes according to the level of decentralization* | | |

## 6. Proposed Model: MPISA

Many decentralized computation and storage solutions use different technologies and are used in different domains. Just like connecting different communication networks to form the Internet, different solutions can be inter-connected and their services can be associated. Different decentralized solutions generally use different platforms that are suitable for that domain. By way of example, a scenario may require co-working of a PoW-based cryptocurrency and a PoA-based supply chain solution. Hence, there is a need for a unifying platform that will solve the interoperability problem.

We propose a model called MPISA, whose name is a portmanteau of "Multi-Platform Interoperable Scalable Architecture". We aim to show how multiple platforms can be used together and help developers in solving scalability and interoperability issues. The MPISA model is shown in a two blockchain platform scenario in Figure 10. In this scenario, the two blockchain platforms have their own P2P network and a mainchain as the main blockchain. Each platform uses its own sidechain structure for the scalability issues.

Common data such as digital identities or general preferences can be kept in the shared data storage. Such a system will help in preventing unnecessary re-entrance of such data in different parties and also in preventing inconsistencies. Any change of this data will require only one update and will be available to all parties instantaneously. These will decrease the maintenance costs of this data across systems (Houlding, 2019). The data can be kept in a cloud or distributed storage. It can be reached through a decentralized identity management system. Such a platform can be designed to keep user credentials safely. The Dapps will be able to check the user identity through this system. Zero-knowledge proof can also be integrated into this system to ensure the privacy of the parties. Users can keep their credentials on this shared platform without revealing their private data.

The Dapps have relevant APIs to grant access to their associated blockchain platform. Smart contracts are used in the data storage operations. The blockchain only keeps the records of the transactions made, but the associated data is not kept in the ledger. A cloud or distributed storage is used for storing and retrieving data. Data can be reached using the data locations in the ledger records.

The most challenging component of the model is the interoperability platform. The Dapps will be able to reach different blockchains and their associated data using this platform. Sidechains or atomic swaps can be used to enable interoperability between the chains.

However, sidechain solution proposals are mostly proof of concept and experimental (Johnson et.al, 2019). This area and the scalability issues are still open for development.

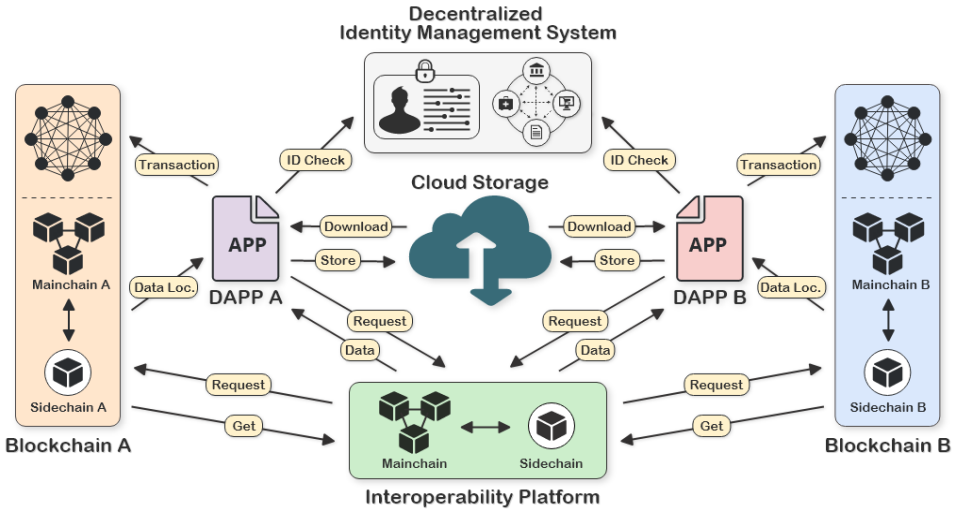

**Figure 10:** Multi-platform interoperable scalable system (MPISA) scenario

## 7. Conclusion

This study aims to describe the decentralized ledger technology and its usage as a data storage to the data scientists and to give a contribution to academia by making the concept easier to understand. Scalability measures in blockchain layers are given in Table 4. Blockchain technology is compared with evolving data solutions in Table 5 and is compared with the relational database in Table 6. We believe that if the data scientists could understand this technology better, they would be able to be a part of the work to solve the challenging issues that come with it.

Blockchain is not a place to store all kinds of different data; it is a registry where the records of transactions are stored. Blockchain is currently the most effective secure way of keeping these records as a ledger, which are distributed in a network. It will help in sharing the data between different parties and enable collaboration. These DLT based solutions ensure the trust without intermediaries. Smart contracts allow autonomous working of the system.

Decentralized systems can be designed to provide common data such as digital identities or general preferences. Such a system will reduce the time needed for the data synchronization across parties and decrease the maintenance costs of this data. We recommend keeping the data in a cloud or distributed storage. Data will be reachable using the data locations in the ledger records.

Nowadays more decentralized application prototypes have become the new focus, and we are now talking about projects which have started to move towards production alongside legacy systems(Brennan et al, 2018). Finance and supply chain are some of the widely used fields of blockchain technology. There are many fields such as health data exchange, know your customer (KYC), smart governance and fraud detection that fit perfectly with the benefits of blockchain technology. We will see blockchain based open global trade digitization platforms in the future, which will ensure secure and instant access to end-to-end supply chain information (Mohan, 2019).

Despite the obvious opportunities of blockchain technology, the challenges of blockchain are still in need of discussion. The most notable architectural challenges are scalability and privacy problems. Other challenges include energy consumption, interoperability, cryptology challenges in the age of quantum computing. These problems should be solved to achieve better implementations in the field. In particular, we should work on scalability problems. Possible solutions should not have an effect on security and decentralization. However, we also believe that some enterprise solutions may also have some centralized parts.

There are evolving data solutions for the decentralized storage challenges. Several solutions such as making blocks bigger, sharding, using more than one chain, and using a decentralized cloud storage network have been proposed. Using solutions like directed acyclic graph (DAG), distributed hash table (DHT) also seem promising.

DAG, DHT, sidechain, gossip protocol and such like can be used to solve the scalability problems of blockchain. Platforms such as Tangle, Hashgraph and Holochain which use these solutions are compared. We think that these derivatives are important for the evolution of decentralized systems. The decentralized solutions promise better TPS rates than the traditional blockchain systems and are likely be preferred in the near future if no security flaws are noticed in their implementations. However, their capabilities have not been tested much and their sustainability has not been tested as long as the known blockchain technologies.

Measures for the privacy of data should be taken. It is a good practice not to store PII on the blockchain and let the user handle his/her own data. Zero Knowledge Proof (ZKP) can be integrated to blockchain systems to ensure privacy.

We proposed a multi-platform interoperable scalable architecture (MPISA) model. We plan to study scalability and interoperability technologies, which can be used to make such a system possible.

In the near future, interoperability will be one of the most important necessities of the business blockchain platforms for benefiting inter-sectoral business solutions with the wide usage of DLT. Sidechain is the potential structure for enhancing the scalability of existing blockchain implementations and a chance for ensuring the interoperability of blockchain technologies. However, it adds more complexity and should be well designed and implemented.

Blockchain immutability may also be argued. Controlled rewriting of blockchain records with chameleon-hashing may be applicable in some cases (Derler et.al, 2019). Different approaches may be appropriate for different implementation areas. Some domains such as the Internet of Things have domain specific characteristics such as frequent data transfers with small content. Domain specific solutions should be developed. IOTA, based on Tangle, is a candidate for a solution; however, it still has many issues which need to be solved.

Blockchain and artificial intelligence (AI) can be used together to complement each other. Revolutionary improvements are possible (Dinh & Thai, 2018; Salah et.al., 2019).

Supercomputers and quantum technology will be further key elements that will shape future implementations of blockchain. Post-quantum blockchain and secure cryptocurrency schemes, which can resist quantum computing attacks, should be studied.

IEEE, ISO and W3C are working on new standards. We need more standardization efforts on blockchain and decentralized systems. We would like to emphasize that blockchain and its derivatives are still evolving. New advanced approaches and better benchmark systems (Gutierrez C., 2019) are being developed. The promises of decentralized implementations are so evident that the challenges should be studied, and more attention should be given to this field.

**Acknowledgements**

We would like to thank MSKU Blockchain Research Group (http://wiki.netseclab.mu.edu.tr/index.php?title=MSKU_BcRG) members (especially Cemal Dak, Şafak Öksüzer, Ahmet Önder Gür) for their contribution to the graphics used in this chapter.

Enis KARAARSLAN, Enis KONACAKLI | 67